\documentclass[12pt]{article}
\def\hi{\noindent \hangindent=1em }
\def\ni{\noindent}
\normalfont

\begin{document}
.
\vskip 0.5in
\Large
\vskip -6pt
\centerline{\bf Including All the Lines}
\vskip 24pt
\centerline{\large Robert L.\ Kurucz}
\vskip 24pt
\footnotesize
\centerline{\it Harvard-Smithsonian Center for Astrophysics, Cambridge, MA 02138, U.S.A.}
\vskip 36pt
\baselineskip=12truept
\hsize 144truemm
\hi
\ni\ \ \ {\bf Abstract.}  
I present a progress report on including all the lines 
in the linelists, including all the lines in the opacities, including 
all the lines in the model atmosphere and spectrum synthesis 
calculations, producing high-resolution, high-signal-to-noise 
atlases that show (not quite) all the lines, so that finally we 
can determine the properties of stars from a few of the lines.

\vskip 12pt
\ni\ \ \ {\bf Keywords:}   atomic data --- molecular data --- atlases ---
sun: atmosphere --- sun: abundances 

\ \ \ \ \ \ \ \ \ \ \ \ \ \
--- stars: atmospheres --- stars: abundances --- supernovae: general

\hi
\ni\ \ \ {\bf PACS:} 31.15.ag \ 31.30.Gs \ 32.30.-r. \ 32.70.-n 
\ 32.70.Cs  33.20.-t \ 33.70.Ca \ 95.30.Ky \ 95.75.Fg

\ni \ \ \ \ \ \ \ \ \ \ \ \ \ \ \
96.60.Fe \ 96.60.Ub \ 97.10.Ex \ 97.10.Tk \ 97.60.Bw  
\vskip 24pt
\hsize=148truemm
\large 
\centerline{\bf INTRODUCTION}
\vskip 12pt
\normalsize
     In 1965 I started collecting and computing atomic and molecular 
line data for computing opacities in model atmospheres and then 
for synthesizing spectra.  I wanted to determine stellar effective 
temperatures, gravities, and abundances.  I still want to.  

     For 23 years I put in more and more lines but I could never get 
a solar model to look right, to reproduce the observed energy 
distribution.

     In 1988 I finally produced enough lines, I thought.  I completed 
a calculation of the first 9 ions of the iron group elements shown 
in Table 1 using my versions of Cowan's atomic structure programs 
(Kurucz 1988) [1].  There were data for 42 million lines that I combined
with data for 1 million lines
from my earlier list for lighter and heavier elements including  
all the data from the literature.  In addition I had computed linelists
for diatomic molecules including 15 million lines of H$_2$, CH, NH, OH, MgH, 
SiH, C$_2$, CN, CO, SiO, and TiO for a total of 58 million lines.

     I then tabulated 2 nm resolution opacity distribution functions
from the line list for temperatures from 2000 to 200000K and for
a range of pressure suitable for stellar atmospheres (Kurucz 1992) [2]. 

     Using the ODFs I computed a theoretical solar model (Kurucz 1992) [3]
with the solar effective temperature and gravity, 
the current solar abundances from Anders and
\eject

\ni \ \ \ \ \ \ {\bf Table 1.}  Iron group lines computed at San Diego 
Supercomputer Center 1988
\vskip 12pt
\footnotesize
\tabskip 8pt
\halign to\hsize%
{\hfil#&#&\hfil#&\hfil#&\hfil#&\hfil#&\hfil#&\hfil#&\hfil#&\hfil#&\hfil#\cr
&  &      I&     II&    III&     IV&      V&     VI&    VII&   VIII&     IX \cr  
&Ca&  48573&   4227&  11740& 113121& 330004& 217929& 125560&  30156&  22803 \cr
&Sc& 191253&  49811&   1578&  16985& 130563& 456400& 227121& 136916&  30587 \cr
&Ti& 867399& 264867&  23742&   5079&  37610& 155919& 356808& 230705& 139356 \cr
&V& 1156790& 925330& 284003&  61630&   8427&  39525& 160652& 443343& 231753 \cr
&Cr& 434773&1304043& 990951& 366851&  73222&  10886&  39668& 164228& 454312 \cr
&Mn& 327741& 878996&1589314&1033926& 450293&  79068&  14024&  39770& 147442 \cr
&Fe& 789176&1264969&1604934&1776984&1008385& 475750&  90250&  14561&  39346 \cr
&Co& 546130&1048188&2198940&1569347&2032402&1089039& 562192&  88976&  15185 \cr
&Ni& 149926& 404556&1309729&1918070&1971819&2211919& 967466& 602486&  79627 \cr 
\cr}
\normalsize
\vskip 12pt

\ni Grevesse (1989) [4], 
mixing length to scale height ratio l/H = 1.25, 
and constant microturbulent velocity 1.5 km/s.  It generally matched 
the observed energy distribution from Neckel and Labs (1984) [5].

     I computed thousands of model atmospheres that I distributed on 
magnetic tapes, then on CDs, and now on my web site, kurucz.harvard.edu. 
They made observers happy.  However, agreement with low resolution 
observations of integrated properties does not imply correctness.

\vskip 12pt
\centerline{\large\bf PROBLEMS}
\vskip 12pt
 
In 1988 
     the abundances were wrong, 
     the microturbulent velocity was wrong,
     the convection was wrong, 
     and the opacities were wrong.

     Since 1965 the Fe abundance has varied by over a factor of 10.
In 1988 the Fe abundance was 1.66 times larger than today.
There was mixing length convection with an exaggerated, constant 
microturbulent velocity.  In the grids of models, the default 
microturbulent velocity  was 2 km/s.  My 1D models still have 
mixing-length convection, but now with a depth-dependent microturbulent
velocity  that scales with the convective velocity.  3D models with 
cellular convection do not have microturbulent velocity at all,
but use the doppler shifts from the convective motions.

     In 1988 the line opacity was underestimated because not enough lines
were included in the linelists.  Table 2 is an outline for the Fe II line 
calculation then.  The higher energy levels that produce series of lines
that merge into ultraviolet continua were not included.  Those levels also
produce huge numbers of weaker lines in the visible and infrared that 
blend and fill in the spaces between the stronger lines.  Also lines of
heavier elements were not systematically included.  And then the additional 
broadening from hyperfine and isotopic splitting was not included.

     In 1988 the opacities were low but were balanced by high 
abundances that made the lines stronger and high microturbulent
velocity that made the lines broader.  
     Now the abundances, the convection, and the opacities are still 
wrong, but they have improved.  I am concentrating on filling out 
the line lists.
 
\vskip 12pt
\centerline{{\bf Table 2.} Fe II in 1988}
\vskip 12pt 
\ni Based on Johansson (1978) [6] 

\ni Even: \ \ 22 configurations; \ 5723 levels; \ 354 known levels;

729 Hamiltonian parameters, all CI; \ 46 free LS parameters; \ dev 142 cm$^{-1}$
\vskip 12pt
\begin{tabular}{llllllllll}
\ni              &                  &d$^{7}$\\
\ni d$^{6}$4s    &d$^{5}$4s$^{2}$   &d$^{6}$4d   &d$^{5}$4s4d   &&&d$^{4}$4s$^{2}$4d &d$^{5}$4p$^{2}$\\
\ni d$^{6}$5s    &d$^{5}$4s5s       &d$^{6}$5d   &d$^{5}$4s5d   &d$^{6}$5g   &              &             &              &\\
\ni d$^{6}$6s    &d$^{5}$4s6s       &d$^{6}$6d   &d$^{5}$4s6d   &d$^{6}$6g   &              &             &              &\\
\ni d$^{6}$7s    &                  &d$^{6}$7d   &              &d$^{6}$7g   &              &             &              &\\
\ni d$^{6}$8s    &                  &d$^{6}$8d   &              &            &              &             &              &\\
\ni d$^{6}$9s    &                  &            &              &            &              &             &              &\\
\end{tabular}
\vskip 12pt
\ni Odd: \ \ 16 configurations; \ 5198 levels; \ 435 known levels;

541 Hamiltonian parameters, all CI; \ 43 free LS parameters; \ dev 135 cm$^{-1}$ 
\vskip 12pt
\begin{tabular}{llllllllll}
 
d$^{6}$4p    &d$^{5}$4s4p       &d$^{6}$4f   &d$^{5}$4s4f &d$^{4}$4s$^{2}$4p &\\
d$^{6}$5p    &d$^{5}$4s5p       &d$^{6}$5f   &\\
d$^{6}$6p    &d$^{5}$4s6p       &d$^{6}$6f   &\\
d$^{6}$7p    &d$^{5}$4s7p       &\\
d$^{6}$8p    &d$^{5}$4s8p       &\\
d$^{6}$9p    &\\
\end{tabular}
\vskip 12pt
 
\ni Total E1 lines saved \ \ \ \ \ \ \ \ \ \ \ \ \ \ \ \ \ \ \   1264969

\ni E1 lines with good wavelengths \ \ \ \ 45815 

\vskip 12pt
\vskip 12pt
\centerline{\large\bf EXAMPLES OF NEW CALCULATIONS}
 
\vskip 12pt
Here I show sample statistics from my new semiempirical calculations for Fe II,
Ni I, and Co I to illustrate how important it is to do the basic physics
well and how much data there are to deal with.  Ni, Co, and Fe are prominent
in supernovas, including both radioactive and stable isotopes.  There is not
space here for the lifetime and gf comparisons.  Generally, low configurations
that have been well studied in the laboratory produce good lifetimes and gf values
while higher configurations that are poorly observed and are strongly mixed are
not well constrained in the least squares fit and necessarily produce poorer
results and large scatter.  My hope is that the predicted energy levels
can help the laboratory spectroscopists to identify more levels and further
constrain the least squares fits.  From my side, I check the computed gf values
in spectrum calculations by comparing to observed spectra.  I adjust the gf
values so that the spectra match.  Then I search for patterns in the adjustments
that suggest corrections in the least squares fits.

     As the new calculations accumulate I will put on my web site 
the output files of the least-squares fits to the energy levels, energy level
tables, with E, J, identification, strongest eigenvector components, lifetime,
A sum, C$_{4}$, C$_{6}$, Land\'e g.  The sums are complete up to the first (n = 10)
energy level not included.  There will be electric dipole, magnetic dipole,
and electric quadrupole line lists.  Radiative, Stark, and van der Waals
damping constants and Land\'e g values are automatically produced for each line.
Branching fractions are also computed.
Hyperfine and isotopic splitting will be included when the data exist but not
automatically.  Eigenvalues are replaced by measured energies so that lines
connecting measured levels have correct wavelengths.  Most of the lines have
uncertain wavelengths because they connect predicted rather than measured
levels.  Laboratory measurements of gf values and lifetimes will be included.
Measured or estimated widths of autoionizing levels will be included when 
available.  The partition function will be tabulated for a range of densities.
 
When computations with the necessary information are available from other
workers, I am happy to use those data instead of repeating the work.

Once the linelist for an ion or molecule is validated it will be 
incorporated into the wavelength sorted linelists on my website for 
computing opacities or detailed spectra.  The web directories are
kurucz.harvard.edu/atoms and /molecules for the details and /linelists
for the completed linelists.

     Table 3 presents line statistics from some of my recent calculations
that show an order of magnitude increase over my earlier work.  Table 4
shows my estimate that my linelists will have several billion atomic and
molecular lines if I can continue my work.
\vskip 12pt
\centerline {\large\bf Fe II}
\vskip 12pt 
Based on Johansson (1978) [6] and on more recent published and unpublished data.
Johansson had data for more than 100 energy levels that I do not yet have.
\vskip 12pt
\ni Even: \ \ 46 configurations; \ 19771 levels; \ 403 known levels;

\ \ 2645 Hamiltonian parameters, all CI; \ 58 free LS parameters; \ dev 56 cm$^{-1}$
\vskip 12pt

\begin{tabular}{llllllllll}
\ni              &                  &d$^{7}$\\
\ni d$^{6}$4s    &d$^{5}$4s$^{2}$   &d$^{6}$4d   &d$^{5}$4s4d   &&&d$^{4}$4s$^{2}$4d &d$^{5}$4p$^{2}$\\
\ni d$^{6}$5s    &d$^{5}$4s5s       &d$^{6}$5d   &d$^{5}$4s5d   &d$^{6}$5g   &d$^{5}$4s5g&d$^{4}$4s$^{2}$5s\\
\ni d$^{6}$6s    &d$^{5}$4s6s       &d$^{6}$6d   &d$^{5}$4s6d   &d$^{6}$6g   &d$^{5}$4s6g\\
\ni d$^{6}$7s    &d$^{5}$4s7s       &d$^{6}$7d   &d$^{5}$4s7d   &d$^{6}$7g   &d$^{5}$4s7g   &d$^{6}$7i    &d$^{5}$4s7i\\
\ni d$^{6}$8s    &d$^{5}$4s8s       &d$^{6}$8d   &d$^{5}$4s8d   &d$^{6}$8g   &d$^{5}$4s8g   &d$^{6}$8i    &d$^{5}$4s8i  &d$^{5}$4s9l\\
\ni d$^{6}$9s    &d$^{5}$4s9s       &d$^{6}$9d   &d$^{5}$4s9d   &d$^{6}$9g   &d$^{5}$4s9g   &d$^{6}$9i    &d$^{5}$4s9i  &d$^{6}$9l\\
\end{tabular}
\vskip 12pt
\ni Odd: \ \ 39 configurations; \ 19652 levels; \ 492 known levels;

\ \ 2996 Hamiltonian parameters, all CI; \ 51 free LS parameters; \ dev 75 cm$^{-1}$
\vskip 12pt

\begin{tabular}{llllllllll}
 
d$^{6}$4p    &d$^{5}$4s4p       &d$^{6}$4f   &d$^{5}$4s4f   &  &&d$^{4}$4s$^{2}$4p &d$^{4}$4s$^{2}$4f\\
d$^{6}$5p    &d$^{5}$4s5p       &d$^{6}$5f   &d$^{5}$4s5f&&&d$^{4}$4s$^{2}$5p \\
d$^{6}$6p    &d$^{5}$4s6p       &d$^{6}$6f   &d$^{5}$4s6f   &d$^{6}$6h   &d$^{5}$4s6h\\
d$^{6}$7p    &d$^{5}$4s7p       &d$^{6}$7f   &d$^{5}$4s7f   &d$^{6}$7h   &d$^{5}$4s7h\\
d$^{6}$8p    &d$^{5}$4s8p       &d$^{6}$8f   &d$^{5}$4s8f   &d$^{6}$8h   &d$^{5}$4s8h   &d$^{6}$8k    &d$^{5}$4s8k\\
d$^{6}$9p    &d$^{5}$4s9p       &d$^{6}$9f   &d$^{5}$4s9f   &d$^{6}$9h   &d$^{5}$4s9h   &d$^{6}$9k    &d$^{5}$4s9k\\
\end{tabular}
\vskip 12pt

\ni Total E1 lines saved \ \ \ \ \ \ \ \ \ \ \ \ \ \ \ \ \ \ \ new / old = 7719063 / 1254969 \ \ \ ratio = 6     

\ni E1 lines with good wavelengths \ \  new / old = \ \ \ 81225 / 45815 \ \ \  ratio = 1.8
\vskip 12pt
\ni Forbidden lines \ \ \ \ \ \ \ \ \ \ \ \ \ \ \ \ \ \ \  even \ \ \ \ \ \ \ \ \ \ \  odd

\ni total M1 lines saved  \ \ \ \ \ \ \ \ 1852641 \ \ \ \ 2468074 

\ni with good wavelengths \ \ \ \ \ 28102 \ \ \ \ \ \ \ \ 41374

\ni between metastable \ \ \ \ \ \ \ \ \ \ \ \ 1180 \ \ \ \ \ \ \ \ \ \ \ \ \ 0 
\vskip 12pt
\ni total E2 lines saved \ \ \ \ \ \ \   10347332 \ \ \ 13179033 

\ni with good wavelengths \ \ \ \ \ 49019 \ \ \ \ \ \ \ \ \  71225 

\ni between metastable \ \ \ \ \ \ \ \ \ \ \ \ 1704 \ \ \ \ \ \ \ \ \ \ \ \ \ \ 0 

\vskip 12pt

\ni isotopic components \ \ \ \ \ $^{54}$Fe \ \ \  $^{56}$Fe \ \ \ \ $^{57}$Fe \ \ \ \ $^{58}$Fe
 
\ni fractional abundances \ \ \  .059  \ \ \ .9172 \ \ \ .021 \ \ \ .0028   
\vskip 12pt 
\ni $^{57}$Fe has not yet been measured because it has hyperfine splitting.
Rosberg, Litz\'en, and Johansson (1993) [7] have measured  $^{56}$Fe--$^{54}$Fe in 9 lines and
$^{58}$Fe--$^{56}$Fe in one line.  I split the computed lines by hand.
\vskip 12pt
\centerline{\large\bf Ni I}
\vskip 12pt
Ni I mostly based on Litz\'en, Brault, and Thorne (1993) [8] with {\bf isotopic splitting}.
\vskip 12pt

\ni Total E1 lines saved \ \ \ \ \ \ \ \ \ \ \ \ \ \ \ \ \ \ \ \ \ new / old = 732160 / 149926 \ \ \ ratio = 4.9

\ni E1 lines with good wavelengths \ \ \ new / old = \ \ \ 9663 / 3949  \ \ \  ratio = 2.4

\vskip 12pt
\begin{tabular}{llllllllllllllllllll}
isotope&$^{56}$Ni     &$^{57}$Ni     &$^{58}$Ni    &$^{59}$Ni   &$^{60}$Ni   &$^{61}$Ni   &$^{62}$Ni   &$^{63}$Ni   &$^{64}$Ni\\
 
fraction&   \ \ .0     &\ \ .0    &.6827  &\ \ .0   &.2790    &.0113     &.0359     &\ \ .0  &.0091 \\
\end{tabular}
\vskip 12pt
\ni There are 5 stable isotopes.  There are measured splittings for 326 lines
from which I determined 131 energy levels relative to the ground.  These
levels are connected by {\bf 11670 isotopic lines}.  Hyperfine splitting was
included for $^{61}$Ni but only 6 levels have been measured which produce 4
lines with 38 components.  A pure isotope laboratory analysis is needed.
Ni I lines are asymmetric from the splitting and they now agree in shape 
with lines in the solar spectrum.
\vfill
\eject
\centerline {\large\bf Co I}
\vskip 12pt
    Co I based on Pickering and Thorne (1996) [9] and on Pickering (1996) [10]
with hyperfine splitting.  
\vskip 12pt

\ni Total E1 lines saved \ \ \ \ \ \ \ \ \ \ \ \ \ \ \ \ \ \ \ \  new / old = 3771908 / 546130 \ \ \ ratio = 6.9

\ni E1 lines with good wavelengths  \ \ \ new / old = \ \  15441 / 9879 \ \ \ ratio = 2.4

\vskip 12pt  
\vskip 12pt
\ni $^{59}$Co is the only stable isotope.  Hyperfine constants have been
measured in 297 levels which produce {\bf 244264 component E1 lines}.  I have
not yet computed the M1 or E2 components.  The new calculation greatly
improves the appearance of the Co I lines in the solar spectrum.
\vskip 12pt
\vskip 12pt
\ni \hskip 120pt {\bf Table 3.} Sample recent calculations               
\vskip 12pt
\footnotesize
\ni\hskip 100pt config \hskip 49pt levels \hskip 86pt E1 lines

\ni \hskip 90pt even \hskip 11pt odd \hskip 20pt even \hskip 20pt odd
\hskip 17pt good wl \hskip 25pt  total \hskip 35pt old

\tabskip 18pt
\halign to\hsize%
{#&#&#&\hfil#&\hfil#&\hfil#&\hfil#&\hfil#&\hfil#&\hfil#\cr
&Fe&I& 61& 50& 18655& 18850& 93508& 6029023& 789176\cr
&Fe&II& 46& 39& 19771& 19652& 85362& 7615097& 1264969\cr
&Fe&III& 49& 41& 19720& 19820& 33982& 9770250& 1604934\cr
&Fe&IV& 61& 54& 13767& 14211& 8408& 14617228& 1776984\cr
&Fe&V& 61& 61& 6560& 7526& 11417& 7785320& 1008385\cr
&Fe&VI& 61& 61& 2094& 2496& 3535& 1386203& 475750\cr
&\cr
&S&I& 61& 61& 2161& 2270& 24722& 225605\cr
&\cr
&Sc&I& 61& 61& 2014& 2318& 15546& 737992& 191253\cr 
&Sc&II& 61& 61& 509& 644& 3436& 116491& 49811\cr
&\cr
&Ti&I& 61& 61& 6628& 7350& 33625& 4754432& 867399\cr
&\cr
&Mn&I& 44& 39& 18343& 19652& 16798& 1481464& 327741\cr
&\cr
&Co&I& 61& 61& 10920& 13085& 15441& 3771900& 546130\cr
&Co&II& 61& 50& 18655& 19364& 23355& 10050728& 1361114\cr
&\cr
&Ni&I& 61& 61& 4303& 5758& 9663& 732160& 149925\cr
&Ni&II& 61& 61& 10270& 11429& 55590& 3645991& 404556\cr
&Ni&III& 61& 50& 18655& 19364& 21251& 11120833& 1309729\cr
&\cr
&Cu&I& 61& 61& 920& 1260& 5720& 28112\cr
&Cu&II& 61& 61& 4303& 5758& 14959& 622985\cr
&Cu&IV& 55& 50& 9563& 17365& 9563& 11857712\cr
&\cr
&Y&I& 61& 61& 1634& 2141& 5393& 59226\cr              
&\cr}
\normalsize
\ \ \ \ \ \ \ Total:  new / old = 100 million / 12 million \ \ \ ratio = 8 
\vskip 12pt
\vskip 12pt
\vfill
\eject
\ni \ \ \ \ \ {\bf Table 4}. Estimated lines in 3d and 4d group sequences (in millions)
\vskip 12pt
\footnotesize
\tabskip 18pt
\halign to\hsize%
{\hfil#&#&\hfil#&\hfil#&\hfil#&\hfil#&\hfil#&\hfil#&\hfil#&\hfil#&\hfil#&\hfil#&\hfil#\cr

&&I&II&III&IV&V&VI&VII&VIII&IX&X&...\cr
&Ca&.05&\cr
&Sc&.7&.05&\cr
&Ti&5&.7&.05&\cr
&V&14&5&.7&.05&\cr
&Cr&10&14&5&.7&.05&\cr
&Mn&1.5&10&14&5&.7&.05&\cr
&Fe&6&7&10&14&5&.7&.05&\cr
&Co&4&10&7&10&14&5&.7&.05&\cr
&Ni&.7&4&10&7&10&14&5&.7&.05&\cr
&Cu&.03&.6&4&10&7&10&14&5&.7&.05&\cr
&Zn&.1&.03&.6&4&10&7&10&14&5&.7&...\cr
&Ga&&.1&.03&.6&4&10&7&10&14&5&...\cr
&Ge&&&.1&.03&.6&4&10&7&10&14&...\cr
&As&&&&.1&.03&.6&4&10&7&10&...\cr
&Se&&&&&.1&.03&.6&4&10&7&...\cr
&Br&&&&&&.1&.03&.6&4&10&...\cr
&Kr&&&&&&&.1&.03&.6&4&...\cr
&Rb&&&&&&&&.1&.03&.6&...\cr
&Sr&.05&&&&&&&&.1&.03&...\cr
&Y&.7&.05&&&&&&&&.1&...\cr
&Zr&5&.7&.05&\cr
&Nb&14&5&.7&.05&\cr
&Mo&10&14&5&.7&.05&\cr
&[Tc]&1.5&10&14&5&.7&.05&\cr
&Ru&6&7&10&14&5&.7&.05&\cr
&Rh&4&10&7&10&14&5&.7&.05&\cr
&Pd&.7&4&10&7&10&14&5&.7&.05&\cr
&Ag&.03&.6&4&10&7&10&14&5&.7&.05&\cr
&Cd&.1&.03&.6&4&10&7&10&14&5&.7&...\cr
&In&&.1&.03&.6&4&10&7&10&14&5&...\cr
&Sn&&&.1&.03&.6&4&10&7&10&14&...\cr
&Sb&&&&.1&.03&.6&4&10&7&10&...\cr
&Te&&&&&.1&.03&.6&4&10&7&...\cr
&I&&&&&&.1&.03&.6&4&10&...\cr
&Xe&&&&&&&.1&.03&.6&4&...\cr
&Cs&&&&&&&&.1&.03&.6&...\cr
&Ba&&&&&&&&&.1&.03&...\cr
\cr}
\normalsize
TOTAL 3d $>$ 500 MILLION

TOTAL 4d $>$ 500 MILLION
  
+ LANTHANIDE SEQUENCES $>$ 1000 MILLION

+ ALL THE OTHER ELEMENT SEQUENCES
\vfill
\eject
 
\vskip 12pt
\centerline{\large\bf TiO and H$_2$O}
\vskip 12pt
     These are examples of incorporating data from other researchers.
 
     Schwenke (1998) [11] calculated energy levels for TiO including 
in the Hamiltonian the 20 lowest vibration states of the 13 lowest electronic 
states of TiO (singlets a, b, c, d, f, g, h and triplets X, A, B, C, D, E) 
and their interactions.  He determined parameters by fitting the observed
energies or by computing theoretical values. Using Langhoff's transition
moments [12] Schwenke generated a linelist for J = 0 to 300 for the

\ni isotopomers \ \ \ \ \ \ \ \ \ \ \   $^{46}$Ti$^{16}$O\ \ \  $^{47}$Ti$^{16}$O\ \ \  
$^{48}$Ti$^{16}$O\ \ \  $^{49}$Ti$^{16}$O\ \ \  $^{50}$Ti$^{16}$O
 
\ni fractional abundances \ \              .080\ \ \ \ \ \  .073\ \ \ \ \ \ \  
.738 \ \ \ \ \ \ \ \ .055 \ \ \ \ \ \ \ \ .054

\ni My version has 37,744,499 lines. 

Good analyses and a similar semiempirical treatment are needed 
for CaOH, ScO, VO, YO, ZrO, LaO, etc.  

Partridge and Schwenke (1997) [13] treated H$_2$O semiempirically.
They included isotopomers  H$_2$$^{16}$O,  H$_2$$^{17}$O,  H$_2$$^{18}$O, 
and HD$^{16}$O .
My version has 65,912,356 lines.  I hope to obtain a newer linelist
with a billion lines in the near future.

\vskip 12pt
\centerline {\large\bf COMPUTING OPACITY}
\vskip 12pt
My program DFSYNTHE can compute the LTE opacity spectrum 
     of 1 billion lines,
     at 4 million frequencies,
     for 1000 T-P pairs,
     for a range of Vturb.

Those spectra can be statistically processed into ODF tables as they are 
computed, or treated in some other approximation, or they can be saved 
directly.  Instead of dealing with lines, one can just interpolate 
(and doppler shift) the opacity spectra.
\vskip 12pt
\centerline {\large\bf MODEL ATMOSPHERE PROGRAMS}
\vskip 12pt
     My model atmosphere program ATLAS12 can deal with 1 billion lines 
by sampling.  It preselects into a smaller linelist the lines that are 
relevant for the model.  It defaults to 30000 sampling points but it 
could sample a million.  It can treat arbitrary depth-dependent 
abundances and velocities.

     My program ATLAS9 uses ODFs so it is independent of the number of lines.
\vskip 12pt
\centerline {\large\bf ATLASES AND SPECTRUM SYNTHESIS}
\vskip 12pt 
High-resolution, high-signal-to-noise spectra are needed to test
the line data and the spectrum synthesis programs and to 
determine the stellar parameters.

There are no high quality solar spectra taken above the atmosphere.
There are good quality FTS spectra from 2 to 16 microns taken
by the ATMOS experiment on the space shuttle.

There are no high or good quality solar spectra in the ultraviolet.

There are various good quality solar spectra taken through
the atmosphere.  I have been trying to reduce the FTS spectra
taken by James Brault from Kitt Peak to produce central intensity,
limb intensity, flux, and irradiance atlases.

Color figures for the Kitt Peak Flux Atlas, telluric absorption, 
the Kitt Peak Irradiance Atlas, irradiance in the H band,
and a one-angstrom sample spectrum calculation with the lines
labelled are on my website kurucz.harvard.edu/papers/dimitrifest.

\vskip 12pt
\centerline {\large\bf CONCLUDING REMARKS}
\vskip 12pt 

Comments on spectrum synthesis and abundance analysis:

Abundances are generally determined from blended features 
that must be deconvolved by synthesizing the features
including every significant blending line.

In general, one half the discernible lines are missing 
from the lists of lines with good wavelengths.  

Every line has to be adjusted in wavelength, 
damping constants, and gf value.  

Most lines used in abundance analyses are not suitable. 
Including many lines reduces the accuracy.
\vskip 12pt

We do not know anything with certainty about the sun, except its mass.
\vskip 12pt
Inclusion of heavier elements, 
higher stages of ionization, 
additional molecules, 
and higher energy levels, 
will increase the opacity in stellar atmospheres, stellar envelopes, 
stellar interiors, supernovae, galaxies, and the rest of the universe.

Detailed and more complete linelists will allow more accurate 
interpretation of features in spectra and the determination of 
stellar properties at any level of radiation hydrodynamics from 
elementary approximations to the most sophisticated treatments.

\vskip 12pt
\centerline{\large\bf REFERENCES}
\footnotesize
\vskip 12pt
\ni 1. Kurucz, R.L., in {\it Trans. Internat. Astronomical Union} {\bf XXB}, edited by
M.McNally, Kluwer, 
 
\ni \ \ \ \ Dordrecht, 1988. pp. 168-172
 
\ni 2. Kurucz. R.L., {\it Revista Mexicana de Astronomia y Astrofisica}, 
     {\bf 23}, 181-186 (1992)

\ni 3. Kurucz, R.L., ``Model Atmospheres for Population Synthesis" in 
{\it Stellar Populations of 

\ni \ \ \ \ Galaxies}, edited by B. Barbuy and A. Renzini, Kluwer, Dordrecht, 1992, pp. 225-232.

\ni 4. Anders, E. and Grevesse, N.,  {\it Geochimica et Cosmochimica Acta} {\bf 53}, 197-214 (1989)

\ni 5. Neckel, H. and Labs, D., 1984, {\it Solar Physics} {\bf 90}, 205-258 (1984)

\ni 6. Johansson, S., {\it Physica Scripta} {\bf 18}, 217-265 (1978)

\ni 7. Rosberg, M., Litz\'en, U., and Johansson, S., {\it Monthly Notices
Royal Astron. Soc.} {\bf262}, L1-L5 

\ni \ \ \ \ (1993)
 
\ni 8. Litz\'en, U., Brault, J.W., and Thorne, A.P., {\it Physica Scripta} {\bf 47}, 628-673 (1993)
 
\ni 9. Pickering, J.C. and Thorne, A.P., {\it Astrophys. J. Supp.} {\bf 107}, 761-809 (1996)
 
\ni 10. Pickering, J.C., {\it Astrophys. J. Supp.} {\bf 107}, 811-822 (1996)
 
\ni 11. Schwenke, D.W., {\it Faraday Discussions} {\bf 109}, 321-334 (1998)

\ni 12. Langhoff, S.R., {\it Astrophys. J.} {\bf 481}, 1007-1015 (1997)
 
\ni 13. Partridge, H. and Schwenke, D.W., {\it J. Chem. Phys.} {\bf 106}, 4618-4639 (1997)

\vfill
\end{document}